\documentclass[traditabstract,referee]{aa}

\usepackage{xspace}
\usepackage{natbib}
\usepackage{graphicx}
\usepackage{amsmath}
\usepackage{amssymb}
\usepackage{txfonts}

\newcommand{\xsrc}{\mbox{GX~301--2~}}
\newcommand{\xsrcnos}{\mbox{GX~301--2}}
\newcommand{\wray}{\mbox{Wray~977~}}
\newcommand{\vela}{Vela\,X-1\xspace}
\newcommand{\Msun}{\ensuremath{\mbox{M}_\odot}\xspace}
\newcommand{\Lsun}{\ensuremath{\mbox{L}_\odot}\xspace}
\newcommand{\ca}{\mbox{\ensuremath{\sim}}}
\newcommand{\nh}{\ensuremath{N_{\text{H}}}\xspace}
\newcommand{\xte}{\textsl{RXTE}\xspace}

\begin{document}

\title{Discovery of a Peculiar Dip from \xsrc}

\author{Ersin {G\"o\u{g}\"u\c{s}}\inst{1} \and Ingo~Kreykenbohm\inst{2} \and Tomaso~M.~Belloni\inst{3}}

\offprints{E. G\"o\u{g}\"u\c{s},\\ e-mail: ersing@sabanciuniv.edu}

\institute{Sabanc\i~University, Faculty of Engineering and Natural Sciences, 
Orhanl\i - Tuzla, 34956 \.Istanbul, Turkey \and Dr. Karl Remeis-Sternwarte \& ECAP, 
Universit\"at Erlangen-N\"urnberg, Sternwartstr. 7, 96049 Bamberg, Germany \and 
INAF$-$Osservatorio Astronomico di Brera, Via E. Bianchi 46, I-23807 Merate (LC), Italy}

\date {Received: October 11, 2010 / Accepted: November 16, 2010}

\abstract{We present temporal and spectral properties of a unique X-ray dip in \xsrc as seen with Rossi X-ray Timing Explorer in May 2010. The X-ray pulsation from the source gradually declined prior to the dip, disappears for one spin cycle during the dip and is abruptly restored in the spin cycle immediately after the dip. Moreover, the phase-integrated spectrum of the source becomes softer before and during the dip and it quickly hardens again following the dip. Our findings indicate the fact that the mechanism for pulsations gradually turned off briefly and underlying dim and softer emission likely from the accretion column became observable in the brief absence of high level emission due to wind accretion.}


\maketitle

\section{Introduction}
\label{Sect:intro}

\xsrc (also known as 4U 1123$-$62) is a High Mass X-ray Binary (HMXB) consisting of a neutron star and the early type companion, \wray (BP Cru). The two objects are in an eccentric orbit (e = 0.462) with an orbital period of 41.5 days \citep{koh97a}. The spin period of the neutron star is 686 s \citep{sato86b}. The companion star, \wray is often classified as a B1.2Ia star \citep{koh97a}, while \citet{kaper95a} reclassified it as B1 Iae+ hypergiant at a distance of 5.3\,kpc. The latter classification makes \wray with a luminosity of $1.3\times 10^6\,\Lsun$ and a mass of at least 48\,\Msun one of the most luminous and most massive stars in our galaxy. The very high mass loss rate of $\dot{M} \sim10^{-5}\,\Msun\,\text{yr}^{-1}$ gives rise to a very dense but slow ($v_\infty =400\,\text{km}\,\text{s}^{-1}$) stellar wind.

The neutron star in \xsrc accretes directly from the dense stellar wind, resulting in a highly variable X-ray light curve that is typical for wind accretion. X-ray luminosity changes by a factor of five within a time frame of an hour is common in \xsrc \citep{rothschild87a}. The orbital light curve shows a distinct profile: shortly before periastron, the source exhibits X-ray flaring activity, \citep[the X-ray luminosity increases by a factor of \ca25][]{pravdo95a}. Around orbital phase 0.2, the X-ray intensity reaches a minimum as the source passes through the dense inner regions of the stellar wind until it is almost eclipsed by the stellar companion \citep[see e.g.][and references therein]{pravdo01a,leahy02a}. Following the minimum, the X-ray luminosity increases again with a weak secondary maximum near apastron. This orbital modulation is explained by \citet{leahy02a} by the presence of a spirally formed gas stream flowing from the optical companion to the neutron star. The neutron star intercepts the gas stream twice per orbit: shortly before periastron where the neutron star runs along the path of the gas stream resulting in intense X-ray flaring activity and also near apastrom giving rise to the secondary peak in the orbital light curve.

Earlier spectral studies of \xsrc could describe its X-ray spectrum with a power law modified by an exponential cutoff at higher energies \citep{white83a}. Recent studies, however, were unable to model the spectrum of using simple models, instead they had to employ partial covering models as the spectrum of \xsrc is heavily absorbed \citep[see e.g.][]{labarbera05a,kreykenbohm04a} with \nh ranging from $10^{23}$ up to $2 \times 10^{24}$\,cm$^{-2}$ \citep{labarbera05a}. Around 6.4\,keV a very strong iron fluorescence line with an equivalent width from 200\,eV up to 1800\,eV is present \citep{leahy88a}.  At energies above 20\,keV, the spectrum is further modified by the presence of a cyclotron resonant scattering feature around $\sim37$\,keV \citep{mihara95a}, which was shown to be strongly pulse phase dependent in energy and depth \citep{kreykenbohm04a}.

In this paper we uncover an X-ray dipping behavior that has seen in \xsrc for the first time. In the next section, we introduce the 2010 Rossi X-ray Timing Explorer (RXTE) observations of the source. In \S 3, we present our detailed temporal and spectral investigations of the RXTE observations on 2010 May 28 during which the source exhibited the X-ray dip and finally we discuss our results and possible mechanisms that could result in the observed behavior in \S 4.

\section{Observations and Data Processing}
\label{Sect:observations}

We employ publicly available Rossi X-ray Timing Explorer (RXTE) observations of \xsrc which were performed in 17 pointings between 2010 May 26 and July 22 for a total exposure time of 184.7 ks (Program IDs: 95354 and 95428). Individual pointings had exposures ranging between about 1 and 34 ks. Those observations performed under the program ID 95354 lasted over 10 ks each, providing a set of uniquely long and uninterrupted (by the satellite orbit around Earth) X-ray data of \xsrcnos.

In our study, we used the data collected with the Proportional Counter Array (PCA, Jahoda et al. 2006) only. Typically, one or two proportional counter unit (PCU) out of five available PCUs were operating during the observations. The combination of operational PCUs varied slightly between observations. We employed all available data for the selected observations for timing studies while we used data collected with PCU2 only for time resolved spectral investigations to avoid calibration uncertainties (see \S 3.2). We processed RXTE data using tools of HEASOFT version 6.9 and performed spectral analysis with XSPEC 12.5.1n (Arnaud 1996).

\section{Data Analysis and Results}
\label{Sect:analysis}

\begin{figure}
\includegraphics[width=10.4cm]{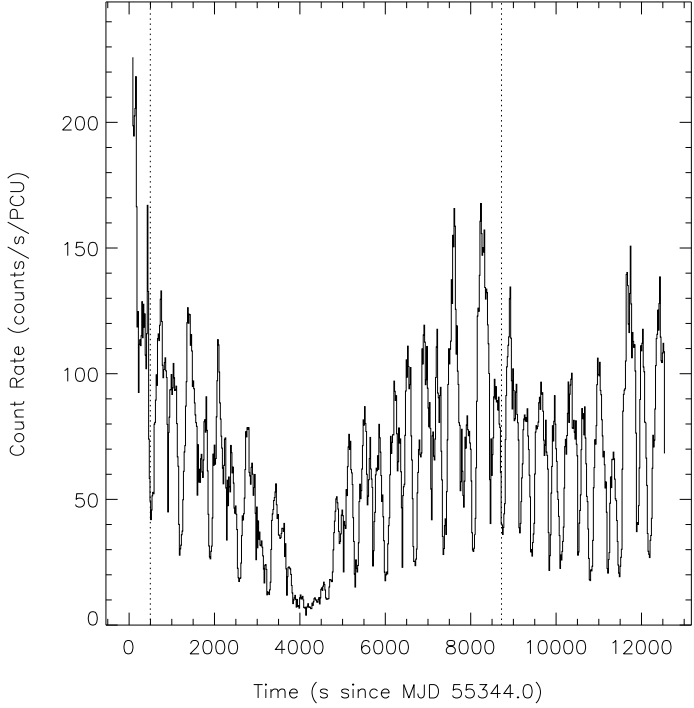}
\caption{Light curve of RXTE/PCA observations of \xsrc on 2010 May 28 (ObsID: 95354-03-03-00) in the 3$-$25 keV range. The vertical dotted lines indicate the time interval of detailed temporal and spectral investigations presented in \S 3.
\label{fig:dipobs}}
\end{figure}

We uncover an X-ray dip in the light curve of \xsrc as seen on 2010 May 28 for the first time. In Figure \ref{fig:dipobs}, we present the 3$-$25 keV light curve of the entire exposure span of about 12.4 ks RXTE pointing of the source. X-ray pulses with a period of 686 s are clearly evident. We find that the X-ray intensity of the source is smoothly decreasing in the episode leading to the dip which lasts about 1000 s (from MJD 55344.044 to 55344.055). X-ray intensity of the source gradually increases after the dip and exhibits typical fluctuations as seen from \xsrc. Our search in the light curves of all 17 RXTE pointings resulted in no additional dipping episode of the source.

We present below the results of our detailed temporal and spectral investigations of RXTE observations on 2010 May 28 (corresponding to the orbital phase of $\sim$0.7 using the ephemeris by Doroshenko et al. [2010]) to reveal the nature of the dipping behavior.

\subsection{Temporal Properties}

\begin{figure}
\includegraphics[width=12.4cm]{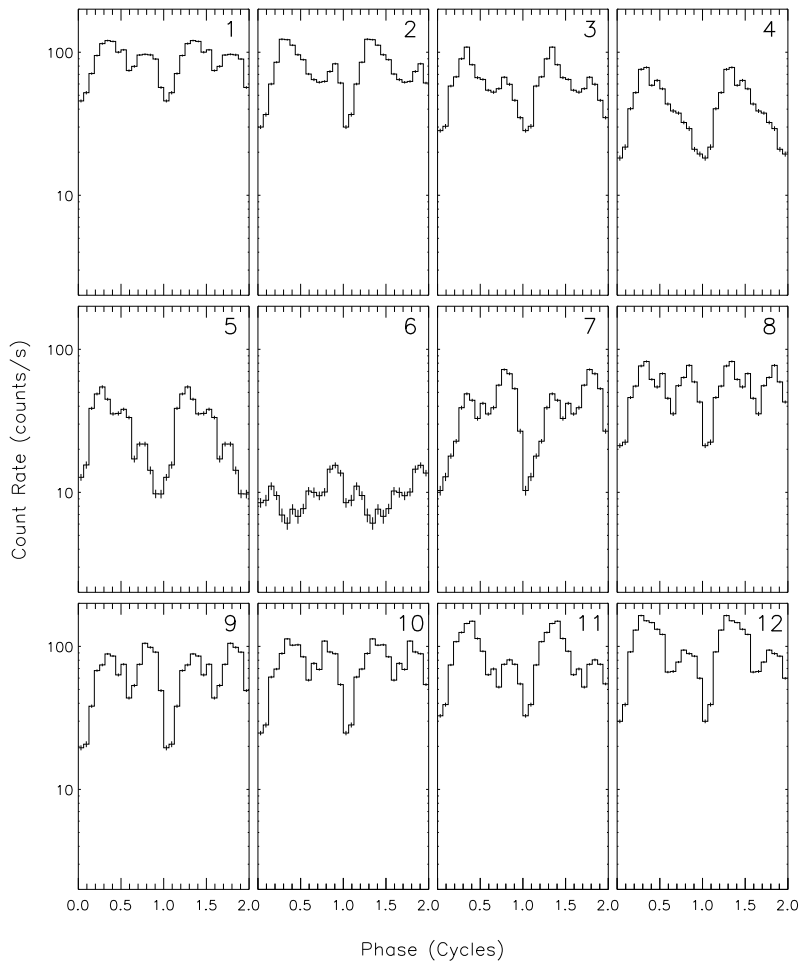}
\vspace{-0.2in}
\caption{Spin phase light curves of \xsrc in the 3$-$25 keV band. Each individual plot is one spin cycle of the pulsar in a continuous time sequence within the time interval indicated in Figure \ref{fig:dipobs}. Notice that Y axes are in logarithmic scale.  
\label{fig:pppanels}}
\end{figure}

In Figure \ref{fig:pppanels}, we show the 3$-$25 keV band light curves of 12 spin cycles of \xsrc in the episode leading to, during, and following the dip. Note that the vertical axis is plotted in logarithmic scale to better identify emission structures with low count rates. The pulse profile of \xsrc is typically characterized by two broad structures; a main pulse and a slightly weaker second peak. We find that in the epoch leading to the dip, while the overall intensity of the X-ray emission decreases, that of the second peak drops faster. Surprisingly, in the dip (cycle number 6), a weak emission structure at the spin phase of the second peak remains while the main pulse structure from \xsrc completely disappears. Following the dip, both emission structures are restored abruptly with the second peak being greater than or equally strong as the first one for four cycles (number 7$-$10). After that, the pulse shape of the source resembles its typical long term appearance again. 

\begin{figure}
\includegraphics[width=10.4cm]{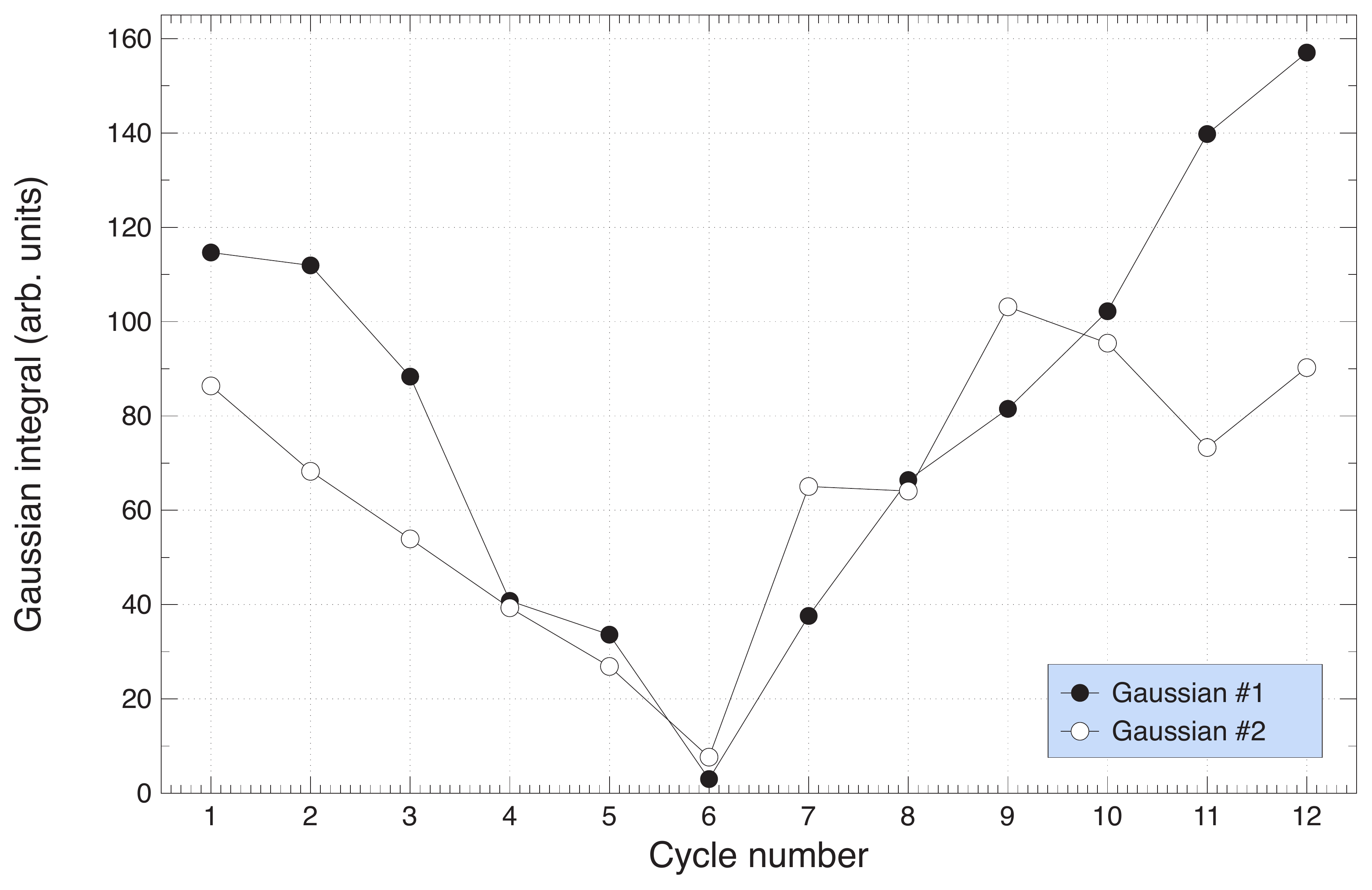}
\caption{Evolution of integrated Gaussian functions 1 and 2 that are used to fit the main pulse and the second peak of the pulse profile.
\label{fig:rmshard}}
\end{figure}

We further investigated evolution of the pulse profile around the dip by means of disentangling the changes in the main pulse and second peak of the pulse profile. For each of the 12 cycles (Figure \ref{fig:pppanels}) we fitted the pulse profile with a model consisting of a constant and two Gaussian components to account for the two peaks. As the constant level was not well constrained as it traded off with the Gaussian normalizations, it was fixed at the rate of 8.4 counts/s/PCU, which is the DC level in the dip. In order to ensure continuity between phase 1 and phase 0, which correspond to the same point, we fitted three versions of each Gaussian shifted by one period ahead and behind. In other words, each pulse was modeled with three Gaussians with a fixed spacing of one period, the same width and normalization. The fits are homogeneous across the cycles. Note however that the Gaussian fits to some of the cycles are not perfect ($\chi^2$ ranging between 8.4 and 49 for 10 degrees of freedom), which is not surprising since the pulse profile of \xsrc is more complex than only two Gaussians can describe. Nevertheless, our main purpose is to evaluate time variation of the two broad pulse structures that can be tracked well with two Gaussians. We then computed the integral of each Gaussian (1 for the main pulse and 2 for the second peak) in the phase range [0,1] which we show in Figure \ref{fig:rmshard} as a function of pulse cycle. It is important to note that both Gaussian integrals gradually decline during the cycles leading to the dip and similarly restored gradually following it.

\subsection{Time Resolved Spectral Properties}

We performed a time resolved spectral analysis in order to uncover the nature of X-ray emission from \xsrc prior to, during, and after the dip. The PCA was operating with two PCUs throughout the time range of interest here. However, only PCU2 was operating in the entire time span while PCU3 was active in the first 5.7 ks and PCU4 was operating in the remainder of the observation. Therefore, we accumulated spectral data collected with PCU2 only to avoid any calibration uncertainties. Given the fact that \xsrc is a bright X-ray source, we could obtain high signal to noise spectra even with one PCU. We accumulated phase integrated source and background spectra for the 12 time intervals shown in Figure \ref{fig:pppanels}. We modeled X-ray spectra in the 3$-$25 keV range, where the sensitivity of PCA instrument is the highest (Jahoda et al. 2006).

\begin{figure}
\includegraphics[width=10.4cm]{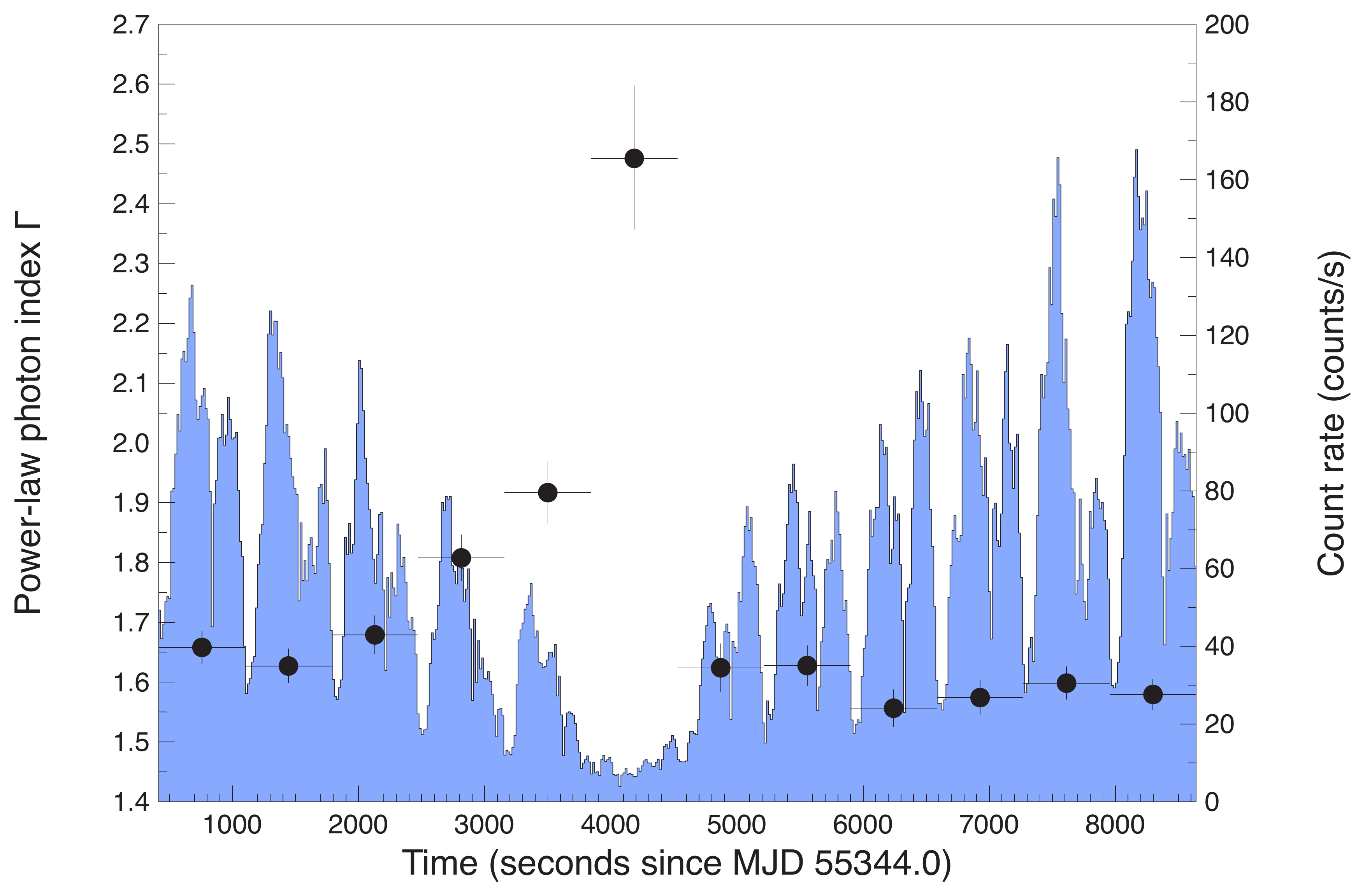}
\caption{Power lax indices that are obtained by spectral modeling (filled circles) and the background subtracted light curve of the source in the 3$-$25 keV range (blue histograms). The horizontal bars indicate one spin cycle (i.e., 686 s) over which each individual spectrum is extracted. The vertical bars are 1$\sigma$ errors in power law indices.
\label{fig:plevol}}
\end{figure}

We first fitted each spectrum individually with an absorbed power law plus a gaussian model. We find that this simple model fits all 12 spectra adequately well ($\chi^2_\nu$ ranging between 0.66 and 1.23). We obtain that the hydrogen column density varies between 1.3$\times$10$^{23}$ and 1.7$\times$10$^{23}$ cm$^{-2}$ but consistent with each other within errors. Similarly, the centroid energy and width of the Gaussian function (that accounts for Fe emission) do not vary across the accumulated spectra. Therefore, we performed a simultaneous fit of all 12 spectra by linking absorption, Gaussian centroid and width parameters but allowing power law index and normalizations of both parameters to vary for all spectra. The resulting best fit yields N$_{\rm H}$ = (1.6$\pm$0.2)$\times$10$^{23}$ cm$^{-2}$ with a fit statistics of $\chi$$^{2}$ / degrees of freedom = 531.4 / 551. Note that the N$_{\rm H}$ we obtain here is consistent with what is determined for the orbital phase interval of 0.62$-$0.65 using BeppoSAX observations (La Barbera et al. 2005). We find that the spectrum of the source becomes softer starting with spin cycle number 4 and the softening trend peaks at the dip (see Figure \ref{fig:plevol}). The spectral energy distribution is restored back to the pre-dip shape immediately after the dip and remains constant (within errors) in the spin cycles investigated here. 

\section{Discussion}
\label{Sect:discussion}

We unveiled intriguing properties of the first observed X-ray dip from \xsrcnos: (i) The X-ray pulsation from the source almost completely disappears for one spin cycle only during the dip and comes back to light in the spin cycle immediately after the dip, (ii) the phase-integrated spectrum of the source becomes softer right before and during the dip and it hardens again in the spin cycle immediately after the dip. Both properties point toward the fact that the mechanism for pulsations gradually turned off briefly, therefore, leading to the observed dip in \xsrcnos.

Although the observed dip in \xsrc is rather unique, other massive wind accreting X-ray binaries also exhibit a similar dipping behavior, in particular, \vela ($P_\text{spin}=283$\,s). In a long term light curve taken with \textsl{INTEGRL}/ISGRI of \vela, \citet{kreykenbohm08a} observed several sudden drops in the count rate: similar to the dip discussed in this work, \vela was also pulsating normally at a flux level of \ca 250\,mCrab, when the pulsations ceased in less than a minute. After about four spin cycles, pulsations resumed back to normal. During these dips in \vela, the source was undetectable in 20\,keV to 40\,keV band by \textsl{INTEGRAL}/ISGRI. 

\begin{figure}[h]
\includegraphics[width=10.4cm]{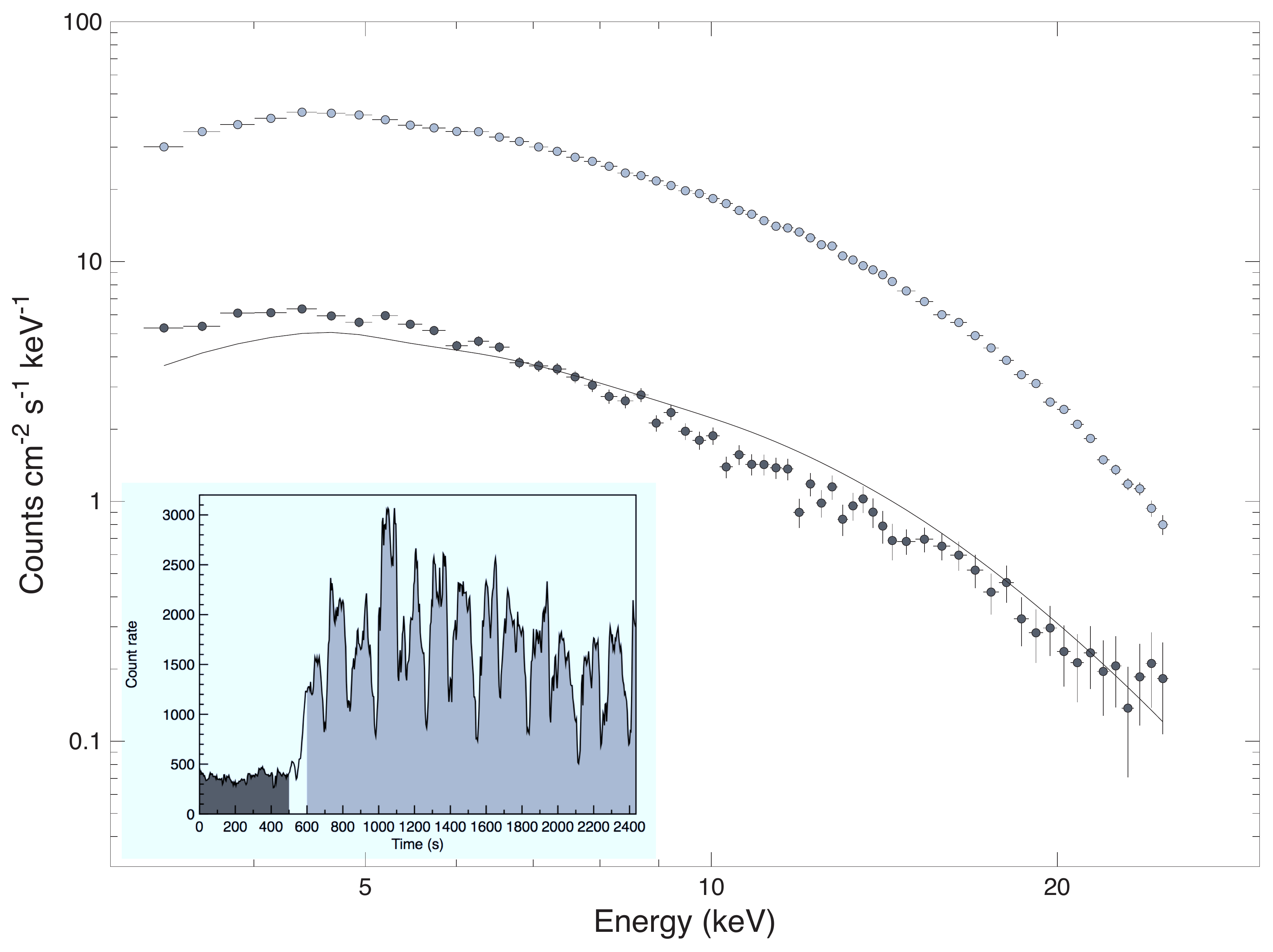}
\caption{Comparison of two X-ray spectra of Vela X-1 during the dip (filled circles, extracted from the time segment marked with dark gray in the inset) and in pulsing phase (empty circles, extracted time segment marked with light gray in the inset). The solid line is the rescaled model curve that describes the non-dip spectrum for comparison with the spectrum of the dip.
\label{fig:velax1}}
\end{figure}

Another such interesting dip was observed in a short \xte observation of again \vela \citep{kreykenbohm99a}: at the beginning of the observation, \vela was not pulsating when the flux was at a very low level (37 counts/s/PCU in the 3$-$25 keV compared to 300 counts/s/PCU during normal bright state; see the inset of Figure \ref{fig:velax1}). Our re-analysis of these data (ObsID: 10141-01-03-00) shows that the source is detected in the PCA, but at a very low flux level and indeed no pulsations are present during this period. We extracted X-ray spectra from the time period when the source was in this dim non-pulsating state and from the bright pulsating state later in the same observation (see Figure \ref{fig:velax1}). The spectral change between the two states is already evident when we rescale the spectral model of the bright pulsating state to the flux level of the dim phase spectrum: the spectrum of \vela is significantly softer during the dim non-pulsating phase. Therefore, \xsrc and \vela show the same behavior: they both exhibit short periods of time during which they are very dim, no pulsation is seen, the spectrum is significantly softer and recovery back to pulsing state is abrupt. 

Gradual reduction and disappearance of X-ray pulsation in \xsrc is suggestive of the fact that X-ray emitting regions of the source were blocked by the wind material \citep{kretschmar99a}. However, one would expect spectral hardening during the dip as softer X-ray photons would be absorbed. Since we observe significant spectral softening during the gap, it is unlikely that the dip is caused by stellar material moving into the line of sight. \xsrc is a source that has exhibited rapid spin up episodes \citep{koh97a}. As a result of an extremely rapid spin-up, the infalling material can be expelled via propeller effect. Cui (1997) reported the disappearence of X-ray pulsations in GX 1+4 and GRO J1744$-$28 during low luminosity states and suggested that they are likely caused by the propeller effect. In case of the dip in \xsrcnos, however, it is unlikely due to the propeller effect since the system would likely remain in the propeller regime much longer than one spin cycle. Another possibility is that there could be a decrease and brief cessation of matter flowing onto the magnetic poles of the neutron star. In this scenario, the observed low level emission could be from the underlying polar region -- a much larger area than the size of the accretion column which would be activated by recent inflow of matter and emission of which can easily be blanketed much brighter emission of accretion. It is still not straightforward to accommodate in this picture why the complete disappearence of the pulse lasts only one spin cycle.

\begin{acknowledgements}
The authors acknowledge EU FP6 Transfer of Knowledge Project Astrophysics of Neutron Stars (MTKD-CT-2006-042722). TMB acknowledges support from ASI grant I/088/06/0.
\end{acknowledgements}


\begin{thebibliography}{12}

\bibitem[Arnaud (1996)]{arn96} Arnaud, K. A., 1996, in Astronomical Data Analysis Software and Systems V, eds. Jacoby G. and Barnes J., p17, ASP Conf. Series Vol. 101

\bibitem[Cui (1997)]{cui07} Cui, W. 1997, ApJ, 482, L163

\bibitem[Jahoda et al. (2006)]{jah06} Jahoda, K. et al., 2006, ApJS, 163, 401

\bibitem[Belloni, Psaltis \& van der Klis (2002)]{bpk02} Belloni, T., Psaltis, D. \& van der Klis, M 2002, ApJ, 572, 392

\bibitem[Doroshenko {et~al.}(2010)]{doro10} Doroshenko, V. Santangelo, A, Suleimanov, V. {et~al.} 2010, A\&A, 515, 10

\bibitem[{Kaper {et~al.}(1995)Kaper, Lamers, Ruymaekers, van~den Heuvel, \& Zuiderwijk}]{kaper95a} Kaper, L., Lamers, H. J. G. L.~M., Ruymaekers, E., van~den Heuvel, E.~P., \& Zuiderwijk, E.~J. 1995, A\&A, 300, 446

\bibitem[{Koh {et~al.}(1997)Koh, Bildsten, Chakrabarty, Nelson, Prince, Vaughn, Finger, Wilson, \& Rubin}]{koh97a} Koh, D.~T., Bildsten, L., Chakrabarty, D., {et~al.} 1997, ApJ, 479, 933

\bibitem[Kretschmar et~al. (1999)]{kretschmar99a} Kretschmar, P. et al. 1999, ApL \& C, 38, 157

\bibitem[{{Kreykenbohm} {et~al.}(1999){Kreykenbohm}, {Kretschmar}, {Wilms}, et al.}]{kreykenbohm99a} {Kreykenbohm}, I., {Kretschmar}, P., {Wilms}, J., {et~al.} 1999, A\&A, 341, 141

\bibitem[{{Kreykenbohm} {et~al.}(2004){Kreykenbohm}, {Wilms}, {Coburn}, {Kuster}, {Rothschild}, {Heindl}, {Kretschmar}, \& {Staubert}}]{kreykenbohm04a} {Kreykenbohm}, I., {Wilms}, J., {Coburn}, W., {et~al.} 2004, A\&A, 427, 975

\bibitem[{{Kreykenbohm} {et~al.}(2008){Kreykenbohm}, {Wilms}, {Kretschmar}}]{kreykenbohm08a} {Kreykenbohm}, I., {Wilms}, J., {Kretschmar}, P., {et~al.} 2008, A\&A, 492, 511 

\bibitem[{{La Barbera} {et~al.}(2005){La Barbera}, {Segreto}, {Santangelo}, {Kreykenbohm}, \& {Orlandini}}]{labarbera05a} {La Barbera}, A., {Segreto}, A., {Santangelo}, A., {Kreykenbohm}, I., \& {Orlandini}, M. 2005, A\&A, 438, 617

\bibitem[{Leahy(2002)}]{leahy02a} Leahy, D.~A. 2002, A\&A, 391, 219

\bibitem[{Leahy {et~al.}(1988)Leahy, Matsuoka, Kawai, Koyama, \& Makino}]{leahy88a} Leahy, D.~A., Matsuoka, M., Kawai, N., Koyama, K., \& Makino, F. 1988, PASJ, 40, 197

\bibitem[{Mihara(1995)}]{mihara95a} Mihara, T. 1995, PhD thesis, RIKEN, Tokyo, Japan

\bibitem[{Pravdo {et~al.}(1995)Pravdo, Day, Angelini, Harmon, Yoshida, \& Saraswat}]{pravdo95a}Pravdo, S.~H., Day, C. S.~R., Angelini, L., {et~al.} 1995, ApJ, 454, 872

\bibitem[{Pravdo \& Ghosh(2001)}]{pravdo01a} Pravdo, S.~H. \& Ghosh, P. 2001, ApJ, 554, 383

\bibitem[{Rothschild \& Soong(1987)}]{rothschild87a}Rothschild, R.~E. \& Soong, Y. 1987, ApJ, 315, 154

\bibitem[{Sato {et~al.}(1986)Sato, Nagase, Kawai, Kelley, Rappaport, \& White}]{sato86b}Sato, N., Nagase, F., Kawai, N., {et~al.} 1986, ApJ, 304, 241

\bibitem[{White {et~al.}(1983)White, Swank, \& Holt}]{white83a}White, N.~E., Swank, J.~H., \& Holt, S.~S. 1983, ApJ, 270, 711

\end{thebibliography}
\end{document}